\documentclass[aps,preprint,showpacs,preprintnumbers,showkeys,eqsecnum,amsmath,amssymb]{revtex4}

\usepackage{graphics}
\usepackage{graphicx}
\usepackage{txfonts}
\usepackage{dcolumn}
\usepackage{mathrsfs}
\usepackage{bm}
\usepackage{amsmath,amssymb,epsfig,float}

\begin{document}

\title{Holographic Superconductors in $z=3$ Ho\v{r}ava-Lifshitz
gravity without condition of detailed balance}

\author{ Jiliang {Jing}\footnote{Electronic address:
jljing@hunnu.edu.cn}}
\author{ Liancheng Wang}
\author{ Songbai Chen}

\affiliation{ Institute of Physics and
Department of Physics,
Hunan Normal University, Changsha, Hunan 410081, P. R. China \\
and
\\ Key Laboratory of Low Dimensional Quantum Structures and
Quantum Control of Ministry of Education, Hunan Normal University,
Changsha, Hunan 410081, P. R. China}

\vspace*{0.2cm}
\begin{abstract}

We study holographic superconductors in a Ho\v{r}ava-Lifshitz black
hole without the condition of the detailed balance. We show that it
is easier for the scalar hair to form as the parameter of the
detailed balance becomes larger, but harder when the mass of the
scalar field larger. We also find that the ratio of the gap
frequency in conductivity to the critical temperature,
$\omega_{g}/T_c$, almost linear decreases with the increase of the
balance constant.  For $\epsilon= 0$ the ratio reduces to Cai's
result $\omega_g/T_c\approx 13$ found in the Ho\v{r}ava-Lifshitz
black hole with the condition of the detailed balance, while  as
$\epsilon \rightarrow 1$ it tends to Horowitz-Roberts relation
$\omega_g/T_c\approx 8$ obtained in the AdS Schwarzschild black
hole. Our result provides a bridge between the results for the
H\v{o}rava-Lifshitz theory with the condition of the detailed
balance and Einstein's gravity.

\end{abstract}



\maketitle

\section{Introduction}

The AdS/CFT correspondence~\cite{Maldacena,polyakov,Witten} relates
a weak coupling gravity theory in an anti-de Sitter space to a
strong coupling conformal field theory in one less dimensions.
Recently it has been applied to condensed matter physics and in
particular to superconductivity \cite{Gubser:2005ih,GubserPRD78}. In
the pioneering papers Gubser \cite{Gubser:2005ih,GubserPRD78}
suggested that near the horizon of a charged black hole there is in
operation a geometrical mechanism parametrized by a charged scalar
field of breaking a local $U(1)$ gauge symmetry. Then, the
gravitational dual of the transition from normal to superconducting
states in the boundary theory was constructed. This dual consists of
a system with a black hole and a charged scalar field, in which the
black hole admits scalar hair at temperature lower than a critical
temperature, but does not possess scalar hair at higher
temperatures~\cite{HartnollPRL101}. In this system a scalar
condensate can take place through the coupling of the scalar field
with the Maxwell field of the background. Much attention has been
focused on the application of AdS/CFT correspondence to condensed
matter physics since then
\cite{HartnollJHEP12,HorowitzPRD78,Nakano-Wen,Amado,
Koutsoumbas,Maeda79,Sonner,HartnollRev,HerzogRev,PW,
Ammon:2008fc,Gubser:2009qm,CJ0}.

H\v{o}rava \cite{ho1,ho2} proposed a new class of quantum gravity.
The key property of this theory is the three dimensional general
covariance and time re-parameterization invariance. It is this
anisotropic rescaling that makes H\v{o}rava's theory power-counting
renormalizable. Therefore, many authors pay their attention to this
gravity theory and its cosmological and astrophysical applications,
and found many interesting results
\cite{KS,CLS,CO,pia,RG,WW,CY,MK,YSM,Nis,CCO1,CJ1,CJ2,DJ0}. These
investigations imply that there exists the distinct difference
between the H\v{o}rava-Lifshitz theory and Einstein's gravity.

Recently, in order to see what difference will appear for the
holographic superconductivity in the H\v{o}rava-Lifshitz theory,
compared with the case of the relativistic general relativity, Cai
{\it et al.} \cite{Cai-Zhang} studied the phase transition of planar
black holes in the Ho\v{r}ava-Lifshitz gravity with the condition of
the detailed balance in which the metric function is described by
$f(r)=x^2-\sqrt{c_0 x}$. They argued that the holographic
superconductivity is a robust phenomenon associated with asymptotic
AdS black holes. And they also got a relation connecting the gap
frequency in conductivity with the critical temperature, which is
given by $ \frac{\omega_g}{T_c}\approx 13, \label{Cai} $ with the
accuracy more than $93\%$ for a range of scalar masses.

Note that the Ho\v{r}ava-Lifshitz black hole without the condition
of the detailed balance has rich physics \cite{Horava,LMP,Cai1},
i.e., changing the parameter of the detailed balance $\epsilon$ from
$0$ to $1$ it can produce the different  black holes for the
H\v{o}rava-Lifshitz theory and Einstein's gravity. Thus, it seems to
be an interesting topic to consider the effects of the parameter of
the detailed balance on the scalar condensation formation, the
electrical conductivity, and the ratio $\omega_g/T_c $ which
connects the gap frequency in conductivity with the critical
temperature.

The paper is organized as follows. In Sec. 2 we present
black holes with hyperbolic horizons in Ho\v{r}ava-Lifshitz gravity in which the action
without the condition of the detailed balance. In Sec. 3 we explore
the scalar condensation in the Ho\v{r}ava-Lifshitz black hole
background by numerical and analytical approaches. In Sec. 4 we
study the electrical conductivity and find ratio of the
gap frequency in conductivity to the critical temperature. We
summarize and discuss our conclusions in the last section.

\section{  black hole with hyperbolic horizon in $z=3$ Ho\v{r}ava-Lifshitz gravity}

In non-relativistic field theory, space and time have different scalings, which is called anisotropic scaling, $  x^i\rightarrow bx^i,$ $t\rightarrow b^zt,$ $ i=1,2,3,$
where $z$ is called {\it dynamical critical exponent}. In order for a theory to be power counting
renormalizable, the critical exponent has at least $z=3$ in four
spacetime dimensions.  For $z=3$, the action without the
condition of the detailed balance for the Ho\v{r}ava-Lifshitz theory
can be expressed as \cite{Horava,LMP}
\begin{eqnarray}
\label{eq6} I &=& \int dt~d^3x [{\cal L}_0 +(1-\epsilon^2){\cal
L}_1],
\end{eqnarray}
with
\begin{eqnarray}
 {\cal L}_0 &=& \sqrt{g}N \left \{\frac{2}{\kappa^2}
(K_{ij}K^{ij}-\lambda K^2) +\frac{\kappa^2\mu^2 (\Lambda
R-3\Lambda^2)}{8(1-3\lambda)}\right \},  \nonumber \\ \nonumber
 {\cal L}_1  &=& \sqrt{g}N \left \{\frac{\kappa^2\mu^2(1-4\lambda)}{32(1-3\lambda)}R^2
-\frac{\kappa^2}{2\omega^4}\left(C_{ij}-\frac{\mu
\omega^2}{2}R_{ij}\right)
\left(C^{ij}-\frac{\mu\omega^2}{2}R^{ij}\right) \right\}.\\ \nonumber
K_{ij}&=&\frac{1}{2N}(\dot g_{ij}-\nabla_iN_j-\nabla_jN_i),\\ \nonumber
C^{ij}&=&\epsilon^{ikl} \nabla _k \left (R^j_{\ l}-\frac{1}{4}R\delta^j_l\right)
= \epsilon^{ikl}\nabla_k R^j_{\ l} -\frac{1}{4}\epsilon^{ikj}\partial_kR,
\end{eqnarray}
where $\kappa^2$, $\mu$, $\Lambda$, and $\omega$  are constant
parameters,  $\epsilon$ is parameter of the detailed balance
($0<\epsilon\leq1$), $N^i$ is the shift vector, $K_{ij}$  is the
extrinsic curvature and $C_{ij}$ the Cotten tensor. It is
interesting to note that the action (\ref{eq6}) reduces to the
action in Ref. \cite{LMP} if $\epsilon=0$, and it becomes the action
for the Einstein's gravity if $\epsilon =1$.

From the action (\ref{eq6}), Cai {\it et al.}  \cite{Cai1} found a
static black hole with hyperbolic horizon whose horizon has an
arbitrary constant scalar curvature $2k$ with $\lambda=1$. The line
element of the black hole can be expressed as
\begin{equation}
\label{metric}
ds^2 = -N^2(r) dt^2 +\frac{dr^2}{f(r)} +r^2 d\Omega_k^2,
\end{equation}
with
\begin{equation}
\label{N2}
N^2 =f = k +\frac{x^2}{1-\epsilon^2}
-\frac{\sqrt{\epsilon^2 x^4+(1-\epsilon^2)c_0 x}}{1-\epsilon^2},
\end{equation}
where $x=\sqrt{-\Lambda}~r$, $k=-1,~0,~1$, and $c_0=(x_+^4+2 k
x_++(1-\varepsilon^2)k^2)/x_+$ in which $x_+$  is the horizon radius
of the black hole, i.e., the largest root of $f(r)=0$. Comparing
with the standard AdS$_4$ spacetime, we may set
$\frac{-\Lambda}{1+\epsilon}=\frac{1}{L_{AdS}^2}$, where $L_{AdS}$
is the radius of AdS$_4$. The authors in ref. \cite{Cai1} also found
that the solutions has a finite mass $M=\kappa^2 \mu^2 \Omega_k
\sqrt{-\Lambda} c_0/16$. For $\epsilon=0$, the solution goes back to
the solution in Ref. \cite{LMP}. Furthermore, when $\epsilon =1$,
the solution becomes the (A)dS Schwarzschild black hole.

The Hawking temperature of the black hole is
\begin{equation}
\label{Hawking temperature}
T = \frac{\sqrt{-\Lambda}}{8\pi}
\frac{3x_+^4+2kx_+^2-(1-\epsilon^2)k^2}{x_+(x_+^2+(1-\epsilon^2)k)},
\end{equation}
which is always a monotonically increasing function of horizon
radius $x_+$ in the physical regime. This implies that the black
holes with hyperbolic horizons in the Ho\v{r}ava-Lifshitz theory are
thermodynamically stable.

\section{ Scalar condensation in Ho\v{r}ava-Lifshitz black-hole background }

Now, we study the scalar condensation in the Ho\v{r}ava-Lifshitz
gravity. In the background of the black hole described by Eq.
(\ref{N2}) with $k=0$, i.e.,
 \begin{eqnarray}\label{tp}
N^2 =f = \frac{x^2}{1-\epsilon^2}
-\frac{\sqrt{\epsilon^2 x^4+(1-\epsilon^2)x x_+^3}}{1-\epsilon^2},
\end{eqnarray}
we consider a Maxwell field and a charged complex scalar field with the action
\begin{eqnarray}\label{System}
S=\int d^{4}x\sqrt{-g}\left[
-\frac{1}{4}F_{\mu\nu}F^{\mu\nu}-|\nabla\psi - iA\psi|^{2}
-V(|\psi|) \right],
\end{eqnarray}
where $F^{\mu\nu}$ is the Maxwell field strength $F=dA$ and $\psi$
is the complex scalar field with the potential $V=m^2|\psi|^2$. We
focus our attention on the case that these fields are weakly coupled
to gravity, i.e., they do not backreact on the metric of the
spacetime. Thus, we can take the ansatz
\begin{eqnarray}
  A_{\mu}&=&(\phi(r),0,0,0),\nonumber \\ \psi&=&\psi(r).
\end{eqnarray}
This ansatz implies that the phase factor of the complex scalar
field is a constant. Therefore, we may take $\psi$ to be real. In
the background of the black hole  described by Eqs. (\ref{metric}) and
(\ref{tp}), the equations of the scalar field $\psi(r)$ and the
scalar potential $\phi(r)$ are given by
\begin{eqnarray}
&&\psi^{\prime\prime}+\left(
\frac{f^\prime}{f}+\frac{2}{r}\right)\psi^\prime
+\left(\frac{\phi^2}{f^2}-\frac{m^2}{f}\right)\psi=0, \label{Psi}
\\
&&\phi^{\prime\prime}+\frac{2}{r}\phi^\prime-\frac{2\psi^2}{f}\phi=0~,
\label{Phi}
\end{eqnarray}
where  a prime denotes the derivative  with respect to $r$.

At the event horizon $r=r_+$, we must have
\begin{eqnarray}
 \psi(r_{+})&=&-\frac{3 \psi^\prime(r_{+})}{2 m^{2} L^2},\nonumber \\  \phi(r_{+})&=&0,
\end{eqnarray}
because their norms are required to be finite, where
$L^2=L_{AdS}^2/(1+\epsilon)$ . And at the asymptotic region
($r\rightarrow\infty$), the solutions behave like
\begin{eqnarray}
\psi&=&\frac{\psi_{-}}{r^{\lambda_{-}}}+\frac{\psi_{+}}{r^{\lambda_{+}}}\,,\nonumber \\
\phi&=&\mu-\frac{\rho}{r}\,, \label{infinity}
\end{eqnarray}
with
\begin{eqnarray}
\lambda_\pm=\frac{1}{2}\Big{[}3\pm\sqrt{9+4m^{2}L_{AdS}^2}~\Big{]}\,,\label{LambdaZF}
\end{eqnarray}
where $\mu$ and $\rho$ are interpreted as the chemical potential and
charge density in the dual field theory, respectively. Because the
boundary is a (2+1)-dimensional field theory, $\mu$ is of mass
dimension one and $\rho$ is of mass dimension two. We can read off
the expectation values of operator $\mathcal{O}$ dual to the field
$\psi$. From Ref. \cite{kw}, we know that for $\psi$, both of these
falloffs are normalizable, and in order to keep the theory stable,
we should either impose
 \begin{eqnarray} \psi{_{-}}=0,\quad \text{and} \quad \langle
 \mathcal{O}_{+}\rangle=\psi{_{+}},\end{eqnarray}
or
 \begin{eqnarray} \psi{_{+}}=0,\quad \text{and} \quad \langle
 \mathcal{O}_{_{-}}\rangle=\psi{_{-}}.\end{eqnarray}
Note that the dimension of temperature $T$ is of mass dimension one,
the ratio $T^2/\rho$ is dimensionless. Therefore increasing $\rho$
while $T$  is fixed, is equivalent to decrease $T$ while $\rho$ is
fixed. In our calculation, we find that when $\rho>\rho_c$, the
operator condensate will appear; this means when $T<T_c$ there will
be an operator condensate, that is to say, the superconducting phase
occurs.

In what following we will present a detailed analysis of the condensation of
the operator $\mathcal {O}_{+}$ by numerical and analytical methods, respectively.

\subsection{Numerical analysis of condensation }

The equations (\ref{Psi}) and (\ref{Phi}) can be solved numerically
by doing integration from the horizon out to the infinity with the
boundary  conditions mentioned above.

\begin{figure}[ht]
\includegraphics[scale=0.5]{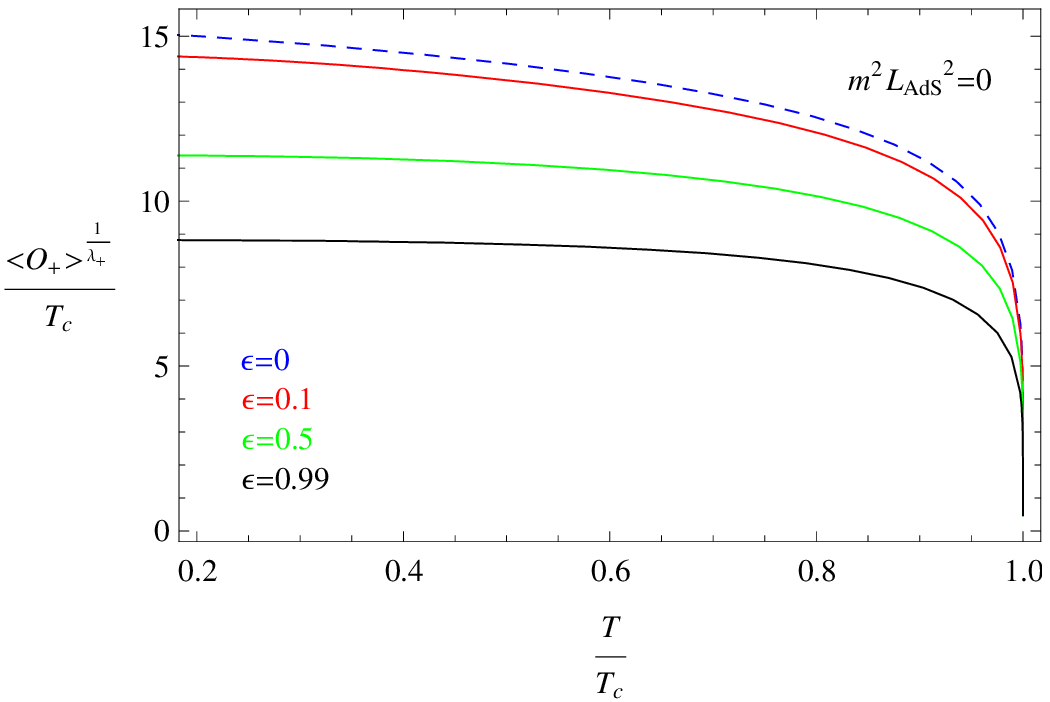}\hspace{0.2cm}%
\includegraphics[scale=0.5]{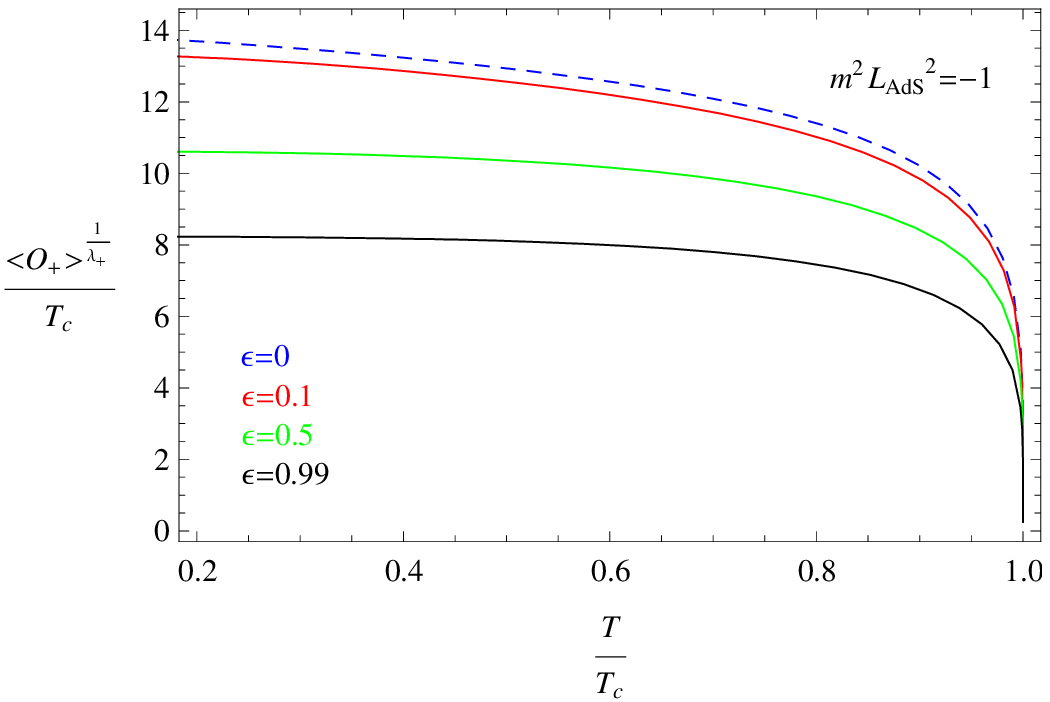}\hspace{0.2cm}%
\includegraphics[scale=0.5]{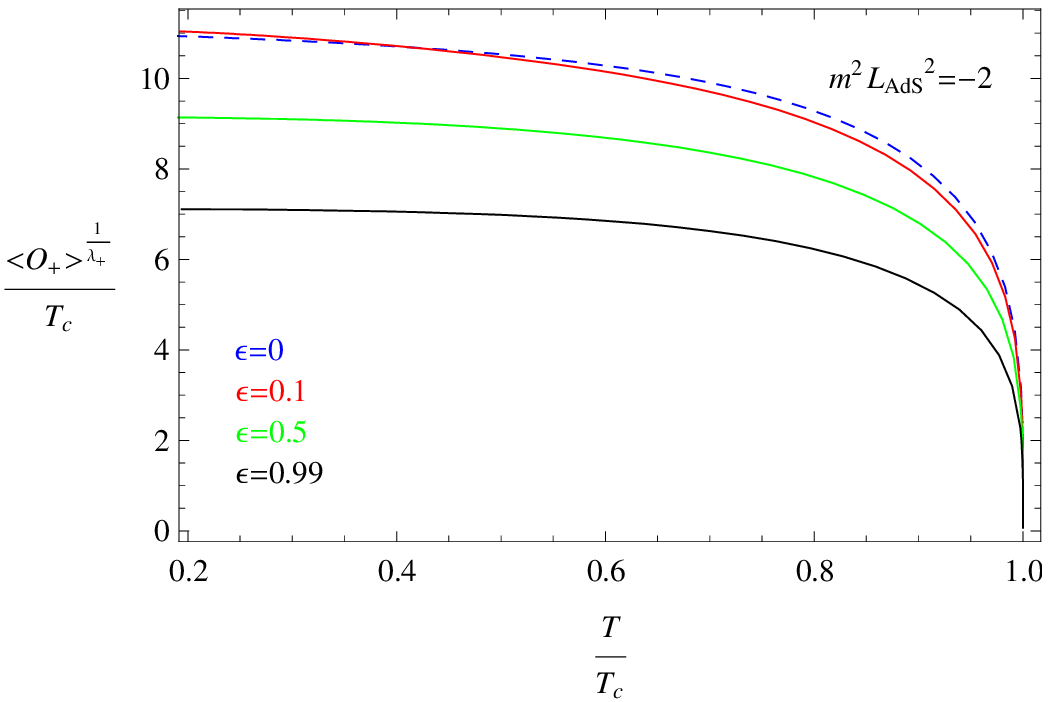}\\ \vspace{0.0cm}
\caption{\label{condensate} (Color online) The condensate as a
function of the temperature with fixed values
$m^{2}L_{AdS}^{2}=0,~-1$ and $-2$. The four lines from top to bottom
correspond to increasing $\epsilon$, i.e., $0$ (blue), $0.1$ (red),
$0.5$ (green) and $0.99$ (black), respectively. It is shown that the
condensation gap becomes smaller as $\epsilon$ increases for the
same mass $m^{2}L_{AdS}^{2}$.}
\end{figure}

\begin{figure}[ht]
\includegraphics[scale=0.75]{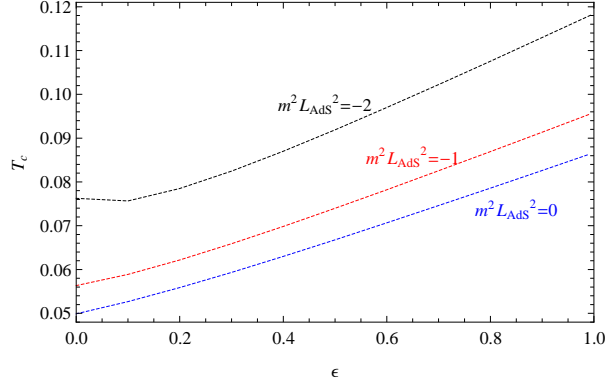}\vspace{0.0cm}
\caption{\label{HLTc} (Color online) The critical temperature as a
function of the balance parameter with fixed values
$m^{2}L_{AdS}^{2}$. The three lines from top to bottom correspond to
$m^{2}L_{AdS}^{2}=-2$ (black), $-1$ (red) and $0$ (blue),
respectively.}
\end{figure}

Changing the value of  $\epsilon$, we present in Fig.
\ref{condensate} the influence of the parameter of the balance on
the condensation with fixed values $ m^2 L_{AdS}^2=0$, $-1$ and
$-2$, and in Fig. \ref{HLTc} the critical temperature  as a function
of the balance parameter with fixed values  $ m^2 L_{AdS}^2=0$, $-1$
and $-2$. We know from the figures that as the parameter of the
detailed balance increases with fixed mass of the scalar field, the
condensation gap becomes smaller, corresponding to larger the
critical temperature, which means that the scalar hair can be formed
easier for the larger $\epsilon$.

\begin{figure}[ht]
\includegraphics[scale=0.5]{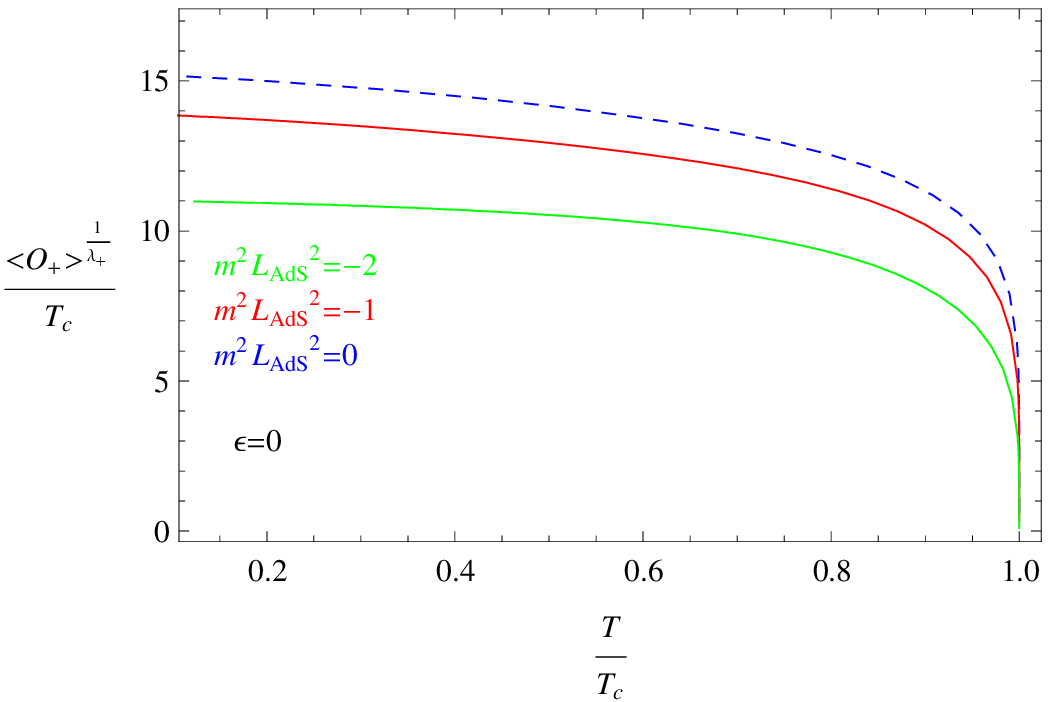}\hspace{0.2cm}%
\includegraphics[scale=0.5]{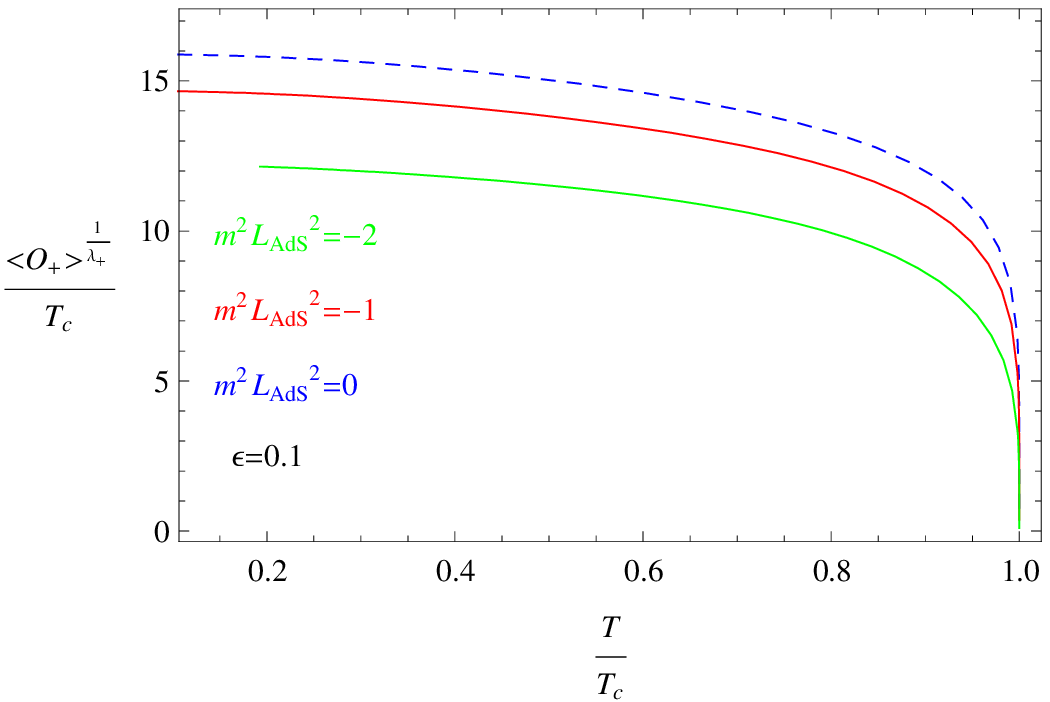}\hspace{0.2cm}%
\includegraphics[scale=0.5]{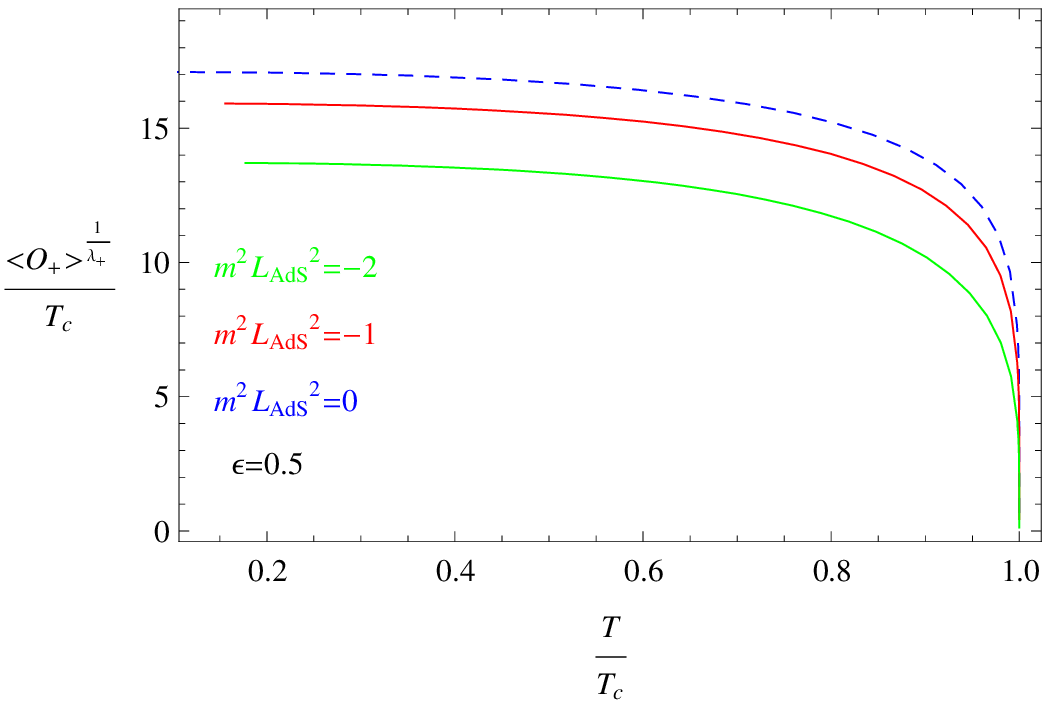}\\ \vspace{0.0cm}
\caption{\label{Cond-eps} (Color online) The condensate as a
function of temperature with fixed values $\epsilon$ for the various
masses of the scalar field. The three lines from bottom to top
correspond to increasing $m^{2}L_{AdS}^{2}$, i.e.,
$m^{2}L_{AdS}^{2}=-2$ (green), $-1$ (red), $0$ (blue), respectively.
It is shown that the condensation gap becomes larger if $m^2$
becomes less negative for fixed $\epsilon $.}
\end{figure}

Altering the value of $m^2 L_{AdS}^2$, we show in Fig.
\ref{Cond-eps} the influence of the mass on the condensation with
fixed values $\epsilon=0$ and $0.1$. It is clear that for the same
$\epsilon $, the condensation gap becomes larger if $m^2$ becomes
less negative, which means that the scalar hair can be formed more
difficult as $m^2$ becomes less negative.

\subsection{Analytical understanding of condensation }

A semi-analytical method can also be applied in understanding the
condensation although the Eqs. (\ref{Psi}) and (\ref{Phi}) are
coupled and nonlinear. The method consists in finding approximate
solutions near the horizon and in the asymptotic AdS space and then
smoothly matching the solutions at an intermediate
point~\cite{Kanti:2002ge}. In particular in Ref. \cite{Gregory} an
analytical expression for the critical temperature was obtained and
the phase transition phenomenon was demonstrated. And the result
obtained in this way is in a good agreement with the numerical
result. In this section we use this analytical approach to study the
condensation of the scalar operator in the Ho\v{r}ava-Lifshitz black
hole, and then compare the result with that obtained by
numerical method.

Take a change $z=r_{+}/r$, Eqs. (\ref{Psi}) and (\ref{Phi}) can be
rewritten as
\begin{eqnarray}
&&\psi^{\prime\prime}+\frac{f^\prime}{f}\psi^\prime
+\frac{r_{+}^2}{z^4}\left(\frac{\phi^2}{f^2}-\frac{m^{2}}{f}\right)
\psi=0~, \label{PsiZ}\\
&&\phi^{\prime\prime}
-\frac{r_{+}^{2}}{z^4}\frac{2\psi^2}{f}\phi=0~, \label{PhiZ}
\end{eqnarray}
where the prime denotes differentiation with $z$. Regularity of the
functions at the horizon $z=1$ requires
\begin{eqnarray}
\psi(1)&=&\frac{3}{2 m^{2}L^{2}}\psi^\prime(1)\,,\nonumber \\
\phi(1)&=&0\,. \label{horizon regularity}
\end{eqnarray}
And near the boundary $z=0$ we have
\begin{eqnarray}
\psi&=&C_{-}z^{\lambda_{-}}+C_{+}z^{\lambda_{+}}\,, \nonumber \\  \phi&=&\mu-\frac{\rho}{r_{+}}z\,. \label{boundary behavior}
\end{eqnarray}
We will  set $C_{-}=0$ and fix $\rho$ in the following discussion.
With the help of the regular horizon boundary condition
(\ref{horizon regularity}), the leading order approximate solutions
near the horizon, $z=1$, for the Eqs. (\ref{PsiZ}) and (\ref{PhiZ})
can be expressed as
\begin{eqnarray}
\psi(z)&=&a  \left\{1+\frac{2 m^{2}L^{2}}{3}(1-z)\right.
\nonumber\\
&&\left.
+\frac{L^2}{36}\left[(3+9\epsilon^2+4 m^2 L^2)m^2
-\frac{ 4 L^{2} b^2}{r_+^{2}}\right](1-z)^2 +
\cdots\right\}, \label{psiTaylor}\\
\phi(z)&=&b\left[(1-z)+\frac{2L^2 a^2 }{3}(1-z)^2 +\cdots\right],
\label{phiTaylor}
\end{eqnarray}
where $a \equiv \psi(1) $ and $ b\equiv -\phi^\prime(1)$ with
$a,~b>0$ which makes $\psi(z)$ and $\phi(z)$ positive near the
horizon. Matching smoothly the solutions (\ref{psiTaylor}),
(\ref{phiTaylor}) with (\ref{boundary behavior})  at an intermediate
point $z_m$ with $0<z_m<1$, we have
\begin{eqnarray}
C_+&=&\frac{6+2m^{2}L^{2}(1-z_{m})}{3 [2z_{m}+\lambda_+(1-z_{m})]z_{m}^{\lambda_{+}-1}}\; a\,, \label{C psi}\\
b&=&\frac{r_{+}}{2L^2} \sqrt{\frac{A} {[\lambda_+-(\lambda_+-2)z_m](1-z_m)}}\equiv \frac{\tilde{b}r_{+}}{L^{2}}~,
\label{b phi}\\
a^2&=&\frac{3}{4 L^2 (1-z_{m})}\left( \frac{\rho}{ b r_+}\right) \left(1-\frac{ b r_+}{\rho}\right)~,
\label{a psi}
\end{eqnarray}
with
\begin{eqnarray}
A&=&4L^4 m^4 (1-z_m)[\lambda_+-(\lambda_+-2)z_m]+36\lambda_+
\nonumber\\ &&+3 L^2 m^2[(1+3\epsilon^2)(\lambda_+-2)z_m^2-2(5+3
\epsilon^2)(\lambda_+-1)z_m+3(3+\epsilon^2)\lambda_+]\,.
\end{eqnarray}

Equation (\ref{b phi}) and the Hawking temperature (\ref{Hawking
temperature}) show that (\ref{a psi}) can be rewritten as
\begin{eqnarray}
a^2=\frac{3}{4(1-z_{m})L^2}\left(\frac{T_c}{T}\right)^{2}
\left[1-\left(\frac{T}{T_c}\right)^{2}\right], \label{rewrite
a psi}
\end{eqnarray}
where $T_c$ is the critical temperature which is defined by
\begin{eqnarray}
T_c=\frac{3}{8\pi L}\sqrt{\frac{\rho}{\tilde{b}}}~.
\label{match Tc}
\end{eqnarray}

According to the AdS/CFT dictionary, we obtain the relation
\begin{eqnarray}
\langle {\cal O_{+}} \rangle\equiv L C_{+} r_{+}^{\lambda_+}
 L^{-2\lambda_{+}}=L C_{+}\left(\frac{8\pi T}{3}\right)^{\lambda_+}.
\end{eqnarray}
Thus, from Eqs. (\ref{C psi}) and (\ref{rewrite a psi}) we find that
the expectation value $\langle {\cal O_{+}} \rangle$ is
\begin{eqnarray}
\label{operator} \frac{\langle {\cal O_{+}}
\rangle^{\frac{1}{\lambda_+}}}{T_c}=\Upsilon~\frac{T}{T_{c}}
\left\{\left(\frac{T_c}{T}\right)^{2}
\left[1-\left(\frac{T}{T_c}\right)^{2}\right]\right\}^{\frac{1}{2\lambda_{+}}},
\end{eqnarray}
where $\Upsilon$ is a constant which is given by
\begin{eqnarray}
\Upsilon=\frac{8\pi}{3}\left\{\sqrt{\frac{3}{4(1-z_{m})}}\frac{6+2 m^2 L^2(1-z_{m})}{ 3 z_{m}^{\lambda_{+}-1}[2 z_m +\lambda_+ (1-z_m)]}\right\}^{\frac{1}{\lambda_{+}}}.
\end{eqnarray}

In table \ref{Tc-D6} we present the critical temperature obtained
analytically by fixing $z_{m}=10/15$ and compare them with the
results obtained by the numerical method. Selecting the appropriate
matching point, we obtain consistent analytic result with that
obtained numerically.

\begin{table}[ht]
\begin{center}
\caption{\label{Tc-D6} The critical temperature $T_{c}$  obtained
by the analytical method (left column) and the numerical method
(right column). The matching point is set as $z_{m}=10/15$.
We have used $\rho=1$ in the calculation.}
\begin{tabular}{c | c | c | c | c | c}
         \hline
$\epsilon$ ~~&0.0 ~~&0.1~~&0.2~~& 0.5~~& 0.9
        \\
        \hline
$m^2L_{AdS}^2=0~$~~~&~$0.0492$~~$0.0499$~~&~$0.0504$~~$0.0502$~~&~
 $0.0515$~~$0.0510$~~&~$0.0544$~~$0.0545$~~&~$0.0577$~~$0.0599$
          \\
$m^2L_{AdS}^2=-1$~~&~$0.0566$~~$0.0563$~~&~$0.0573$~~$0.0589$~~&~
 $0.0580$~~$0.0622$~~&~$0.0602$~~$0.0740$~~&~$0.0631$~~$0.0914$
          \\
 $m^2L_{AdS}^2=-2$~~&~$0.0779$~~$0.0763$~~&~$0.0749$~~$0.0757$~~&~
 $0.0734$~~$0.0785$~~&~$0.0725$~~$0.0919$~~&~$0.0741$~~$0.1130$
          \\
        \hline
\end{tabular}
\end{center}
\end{table}

\section{Electrical Conductivity in Ho\v{r}ava-Lifshitz black-hole background }

In the study of (2+1) and (3+1)-dimensional superconductors,
Horowitz {\it et al.} \cite{HorowitzPRD78} got a universal relation
connecting the gap frequency in conductivity with the critical
temperature $T_c$, which is described by
\begin{eqnarray}
\frac{\omega_g}{T_c}\approx 8,
\end{eqnarray}
with deviations of less than $8\%$. This is roughly twice the BCS
value 3.5 indicating that the holographic superconductors are
strongly coupled. However, the authors in Refs. \cite{PW,Gregory}
found that this relation is not stable in the presence of the
Gauss-Bonnet correction terms. And Cai {\it et al.} \cite{Cai-Zhang}
got a relation
\begin{eqnarray}
\frac{\omega_g}{T_c} \approx 13 \end{eqnarray}
 with the accuracy more than $93\%$ for a
planar Ho\v{r}ava-Lifshitz black hole  with the condition of the
detailed balance. We now examine this result for the
Ho\v{r}ava-Lifshitz gravity considered here.

In order to compute the electrical conductivity, we should study the
electromagnetic perturbation in this Ho\v{r}ava-Lifshitz black hole
background, and then calculate the linear response to the
perturbation. In the probe approximation, the effect of the
perturbation of metric can be ignored. Assuming that the
perturbation of the vector potential is translational symmetric and
has a time dependence as $\delta A_x=A_x(r)e^{-i\omega  t}$, we find
that the Maxwell equation in the Ho\v{r}ava-Lifshitz black hole
background reads
\begin{eqnarray}
A_{x}^{\prime\prime}+\frac{f^\prime}{f}A_{x}^\prime
+\left(\frac{\omega^2}{f^2}-\frac{2\psi^2}{f}\right)A_{x}=0 \; ,
\label{Maxwell Equation}
\end{eqnarray}
where  a prime denotes the derivative  with respect to $r$. An
ingoing wave  boundary condition near the horizon is given by
\begin{eqnarray}
A_{x}(r)\sim f(r)^{-\frac{2 i \omega L^2}{3 r_+}}. \end{eqnarray} In
the asymptotic AdS region ($r\rightarrow\infty$),  the general
behavior should be
\begin{eqnarray}\label{Maxwell boundary}
A_{x}=A^{(0)}+\frac{A^{(1)}}{r} +\cdots~.
\end{eqnarray}
By using AdS/CFT correspondence and the Ohm's law, we know that the
conductivity can be expressed as
  \cite{HorowitzPRD78}
\begin{eqnarray}\label{GBConductivity}
\sigma=\frac{\langle J_x \rangle}{E_x}=-\frac{i\langle J_x
\rangle}{\omega A_x}=\frac{A^{(1)}}{i\omega A^{(0)}} \, .
\end{eqnarray}
\begin{figure}[ht]
\includegraphics[scale=0.375]{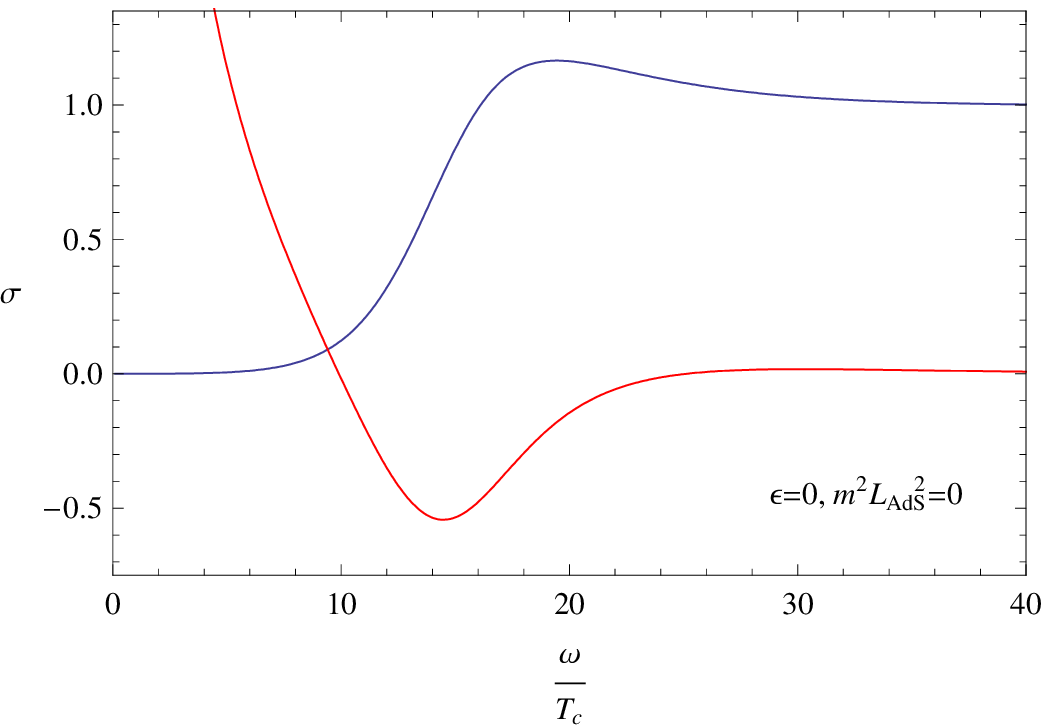}\hspace{0.2cm}%
\includegraphics[scale=0.375]{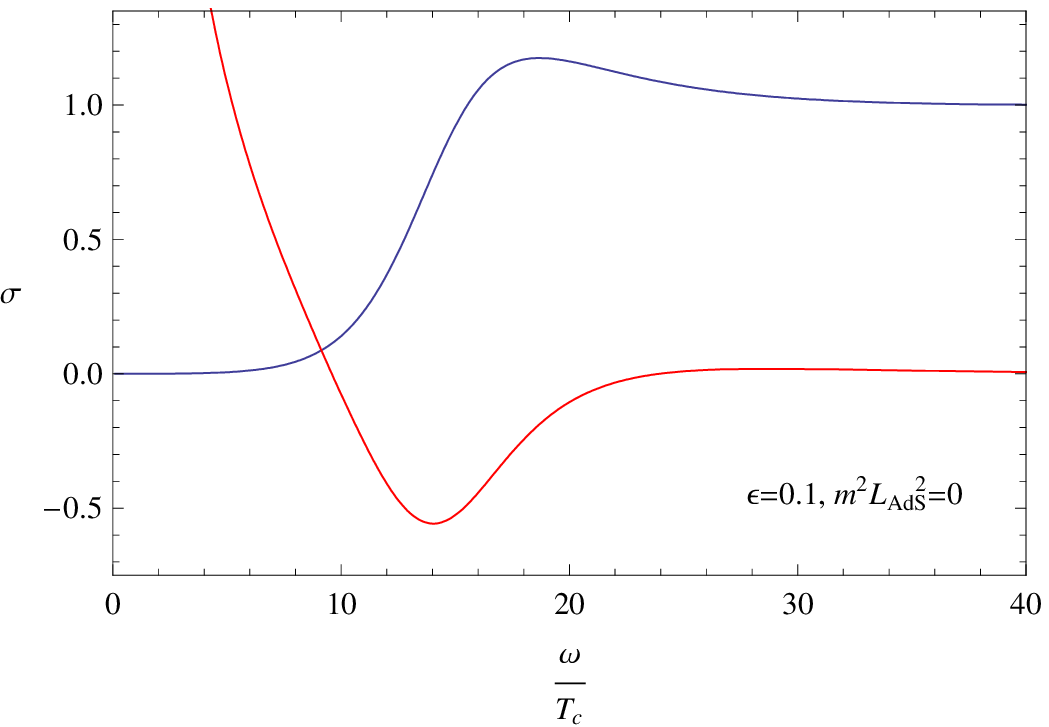}\hspace{0.2cm}%
\includegraphics[scale=0.375]{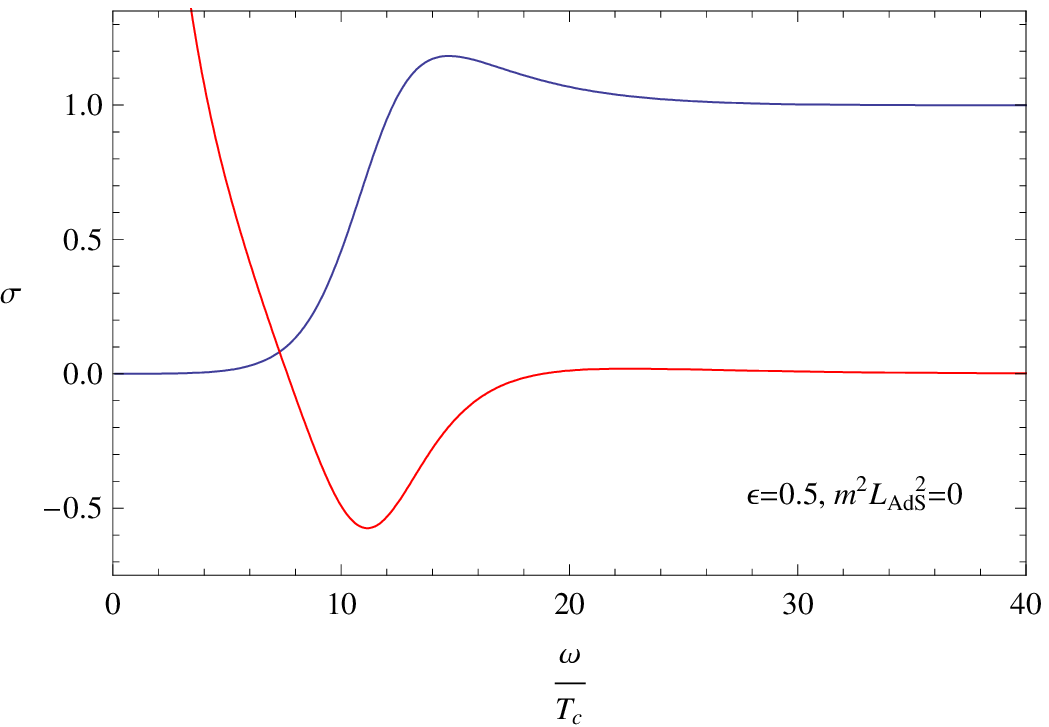}\hspace{0.2cm}%
\includegraphics[scale=0.375]{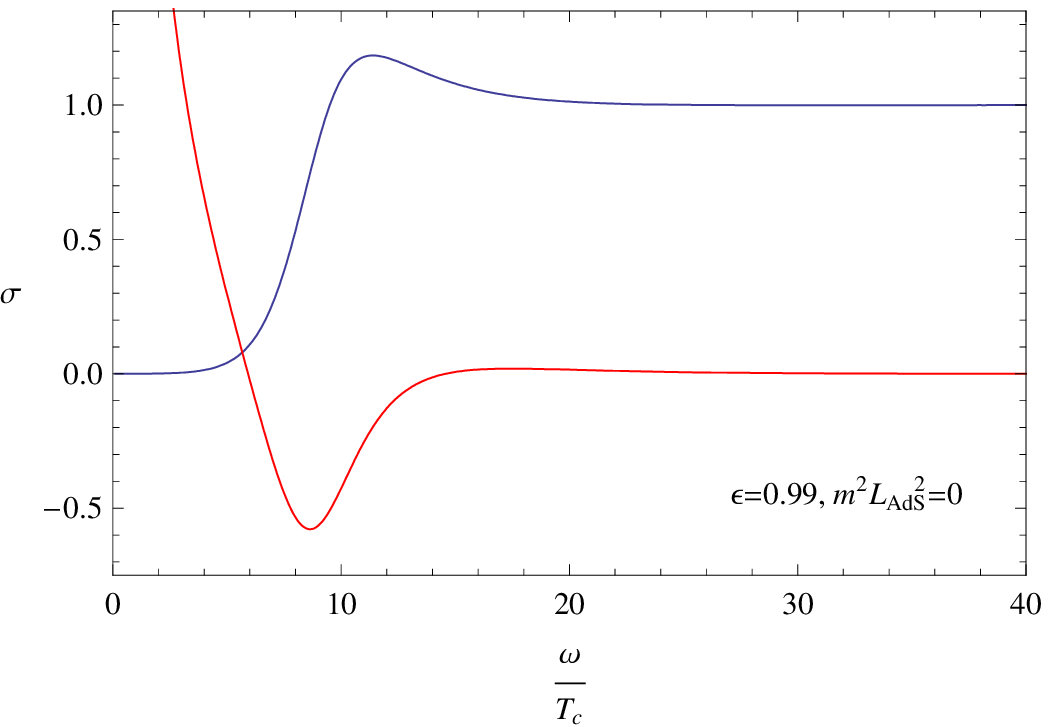}\\ \vspace{0.0cm}
\includegraphics[scale=0.375]{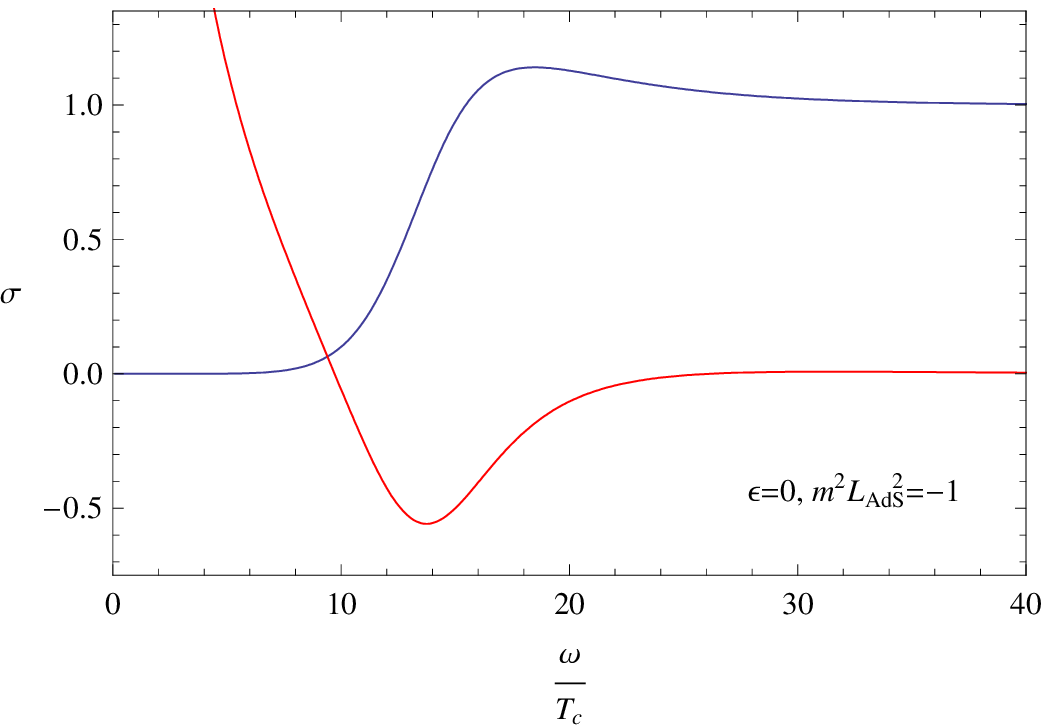}\hspace{0.2cm}%
\includegraphics[scale=0.375]{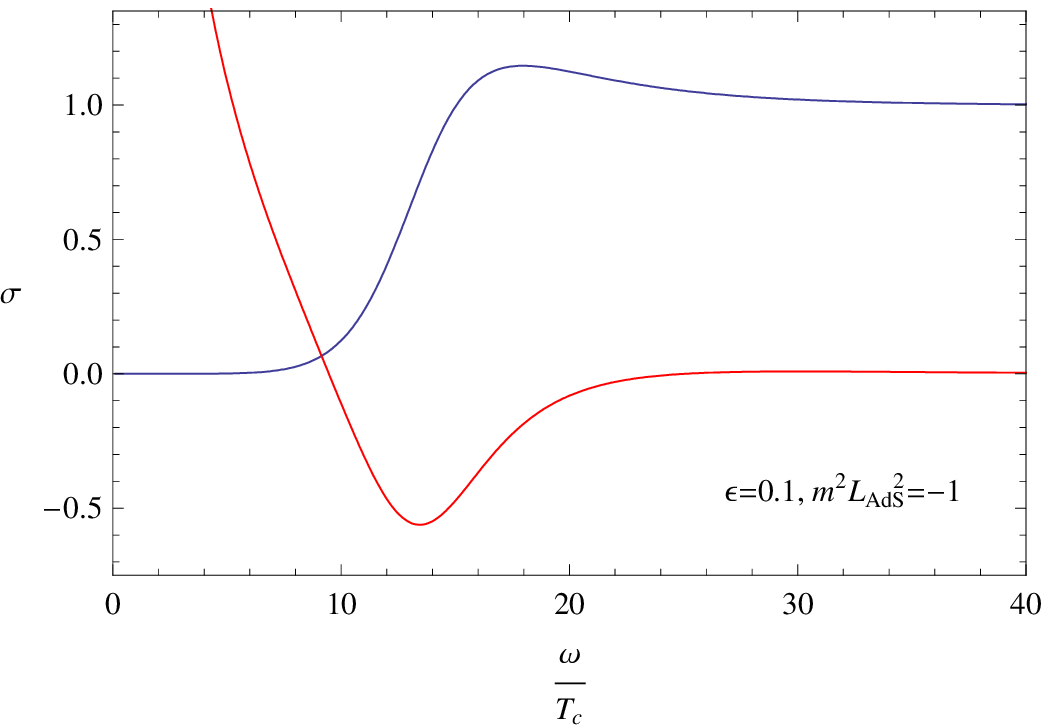}\hspace{0.2cm}%
\includegraphics[scale=0.375]{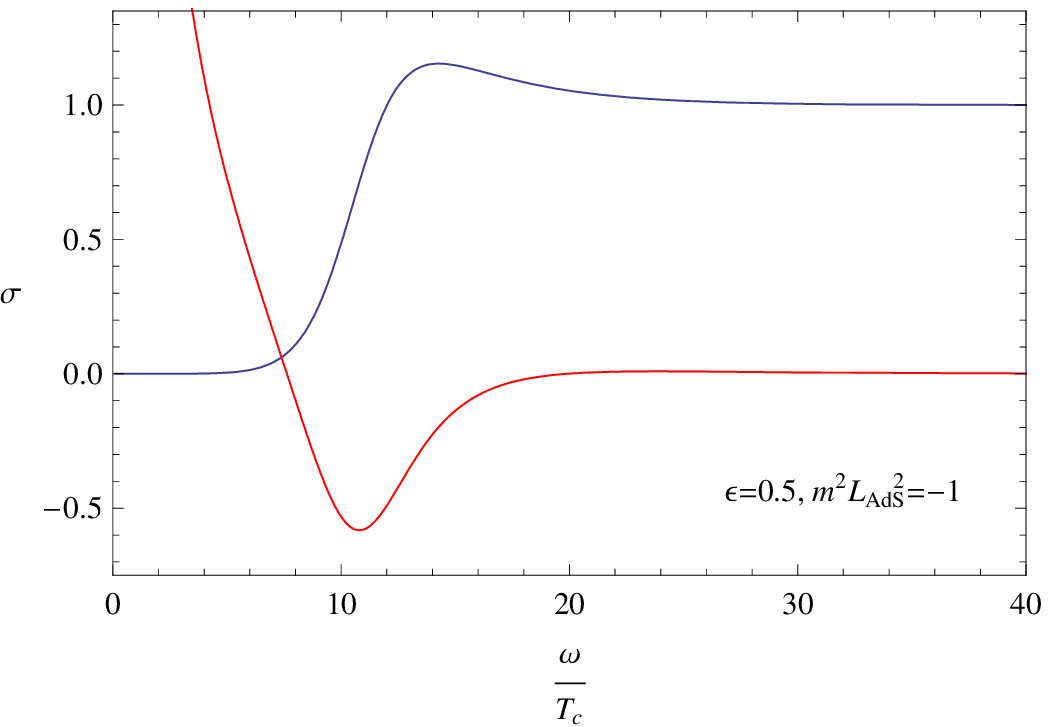}\hspace{0.2cm}%
\includegraphics[scale=0.375]{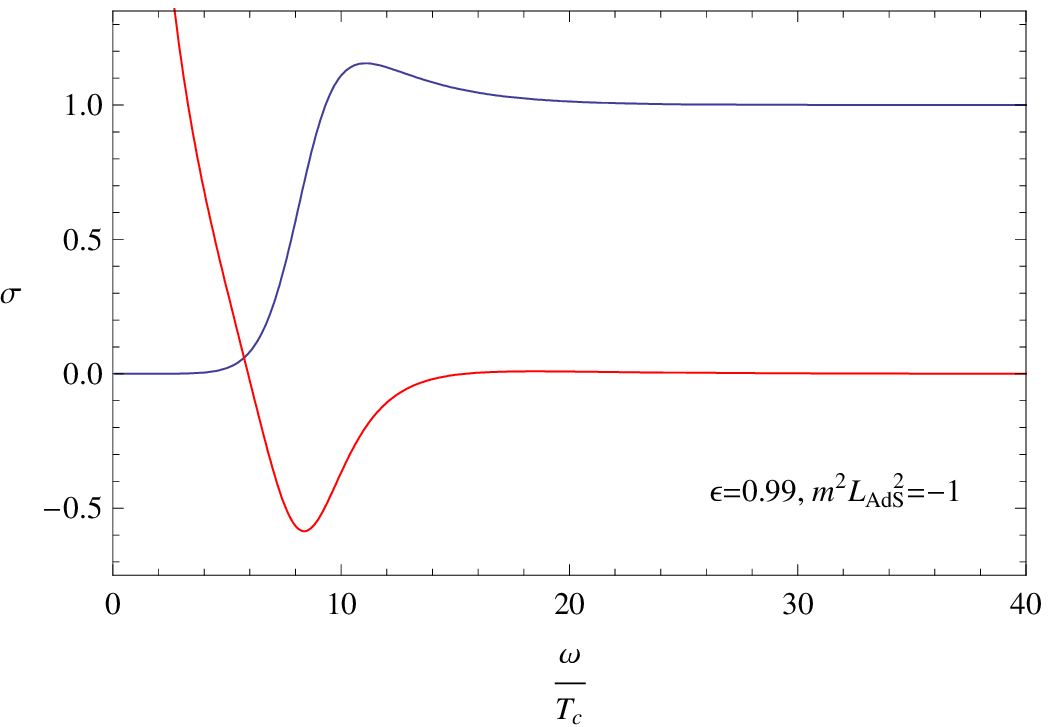}\\ \vspace{0.0cm}
\includegraphics[scale=0.375]{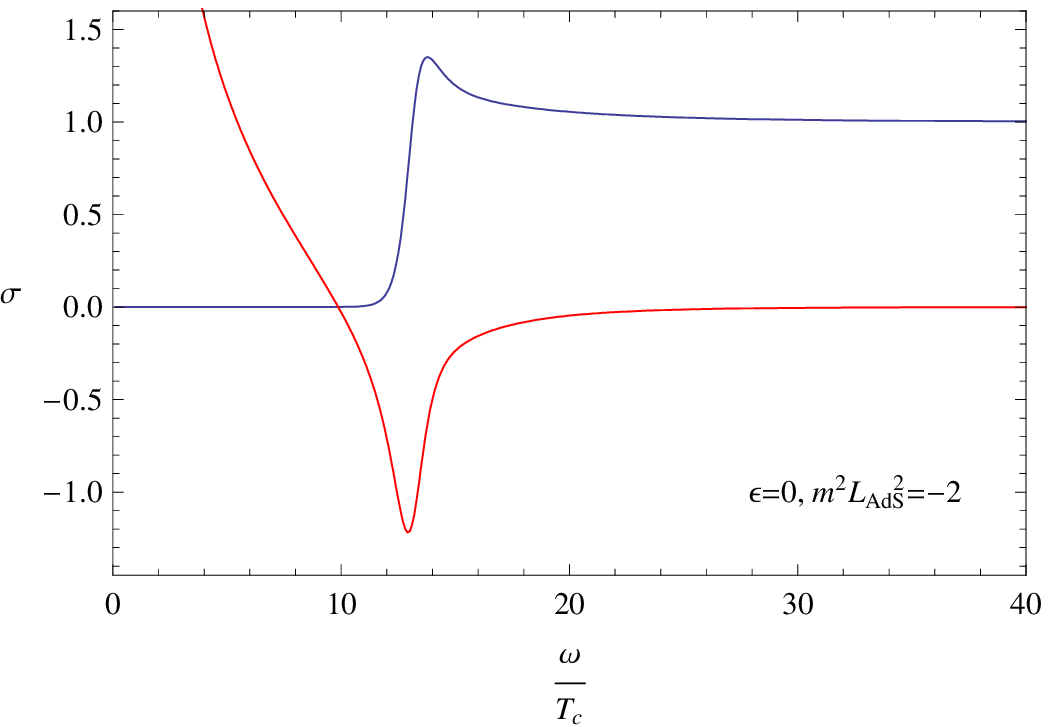}\hspace{0.2cm}%
\includegraphics[scale=0.375]{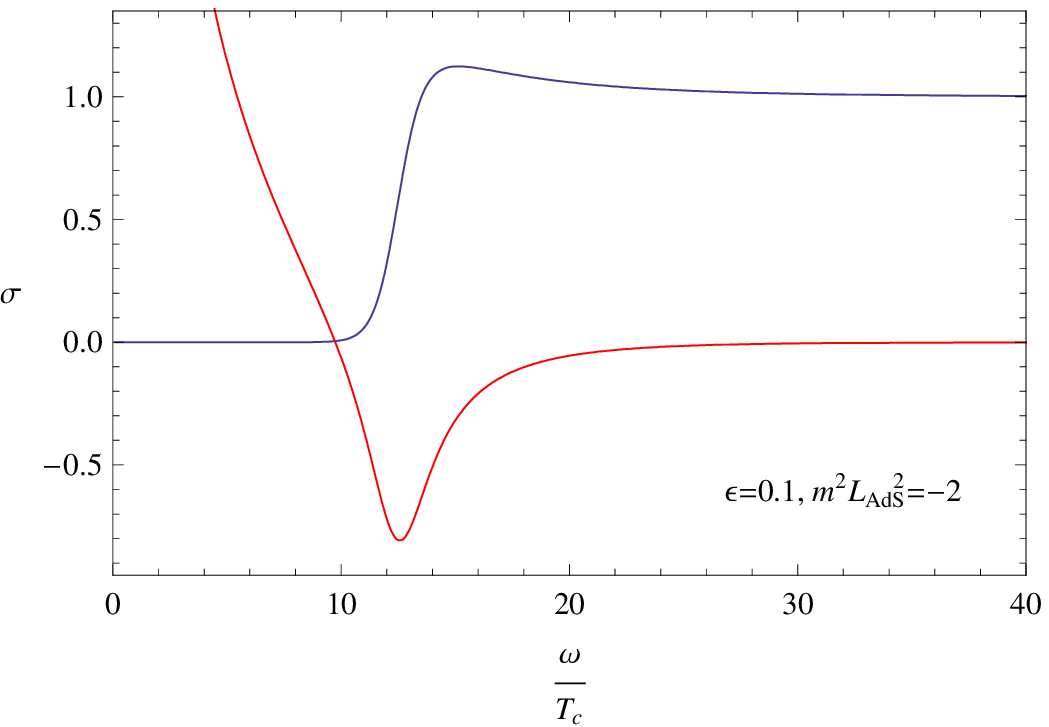}\hspace{0.2cm}%
\includegraphics[scale=0.375]{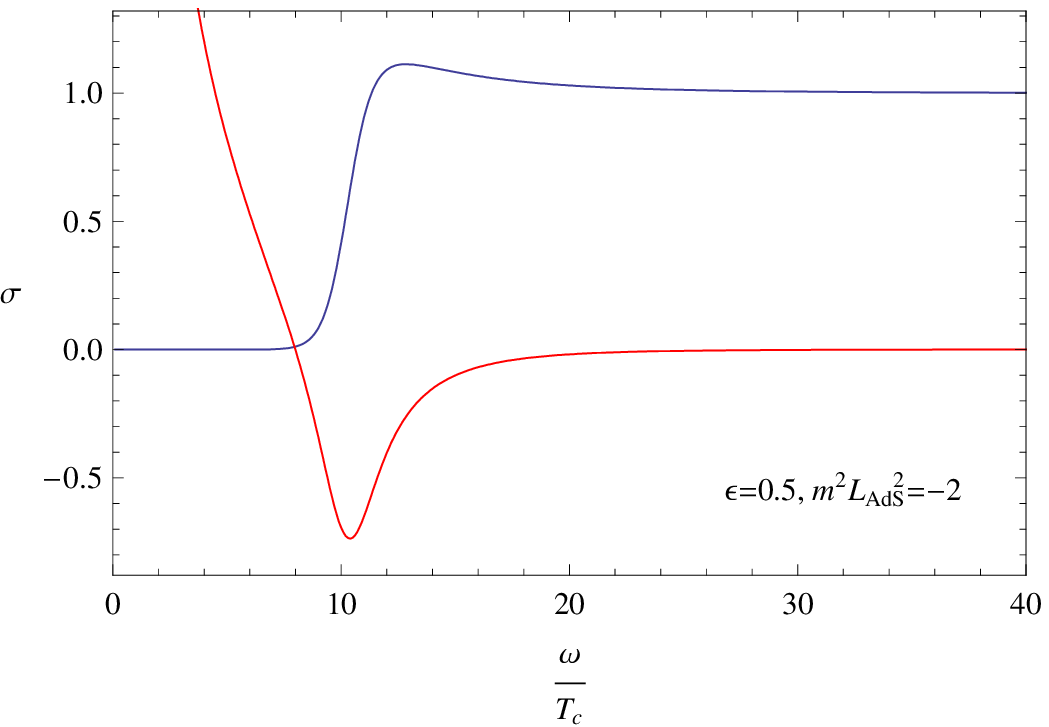}\hspace{0.2cm}%
\includegraphics[scale=0.375]{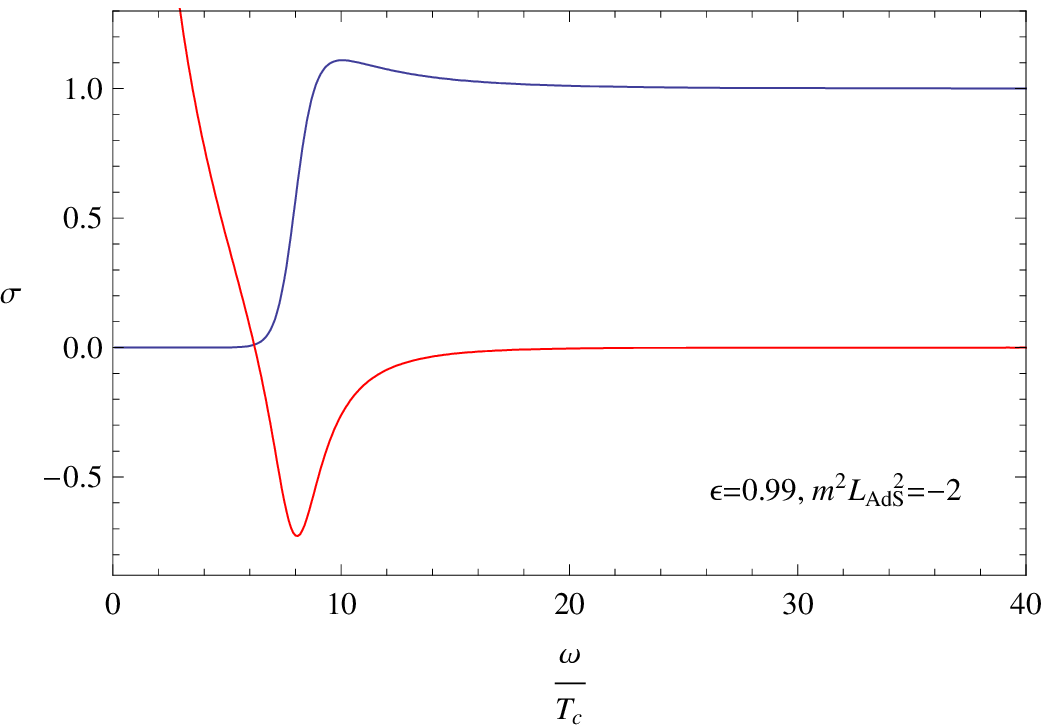}\\ \vspace{0.0cm}
\caption{\label{Conductivity} (Color online) The conductivity of the
superconductors for $\epsilon=0,~0.1,~0.5$ and $0.99$ with
$m^{2}L_{AdS}^{2}=0,~-1$ and $-2$. The solid (blue) line represents
the real part of the conductivity, $Re(\sigma)$, and dashed (red)
line is the imaginary part of the conductivity, $Im(\sigma)$.}
\end{figure}
\begin{figure}[ht]
\includegraphics[scale=0.75]{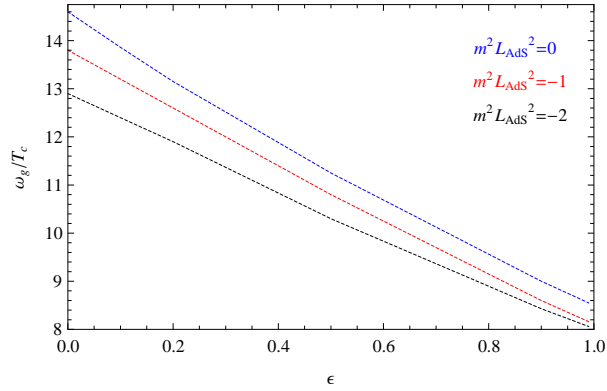}\vspace{0.0cm}
\caption{\label{WgTc} (Color online) The ratio $\omega_{g}/T_{c}$ as
a function of the balance parameter with fixed values
$m^{2}L_{AdS}^{2}$.}
\end{figure}

\begin{table}[ht]
\caption{\label{ConductivityTc} The ratio $\omega_{g}/T_{c}$ for
different values of the constant $\epsilon$ with
$m^{2}L_{AdS}^{2}=0,~-1$ and $-2$. }
\begin{tabular}{|c|c|c|c|c|c|c|}
         \hline
$~~\epsilon~~$ &~~0~~&~~0.1~~&~~0.2~~&~~0.5~~&~~0.9~~&~~0.99~~
          \\
        \hline
~~$m^{2}L_{AdS}^{2}=0$~~ & ~~$14.6$~~ & ~~$13.9$~~ & ~~$13.2$~~&
~~$11.3$~~& ~~$9.0$~~& ~~$8.6$~~
          \\
        \hline
$m^{2}L_{AdS}^{2}=-1$ & $13.8$ & $13.2$ & $12.6$ & $10.8$ & $8.6$ &
$8.2$
          \\
        \hline
$m^{2}L_{AdS}^{2}=-2$ & $12.9$ & $12.4$ & $11.9$ & $10.3$ & $8.4$ &
$8.1$
          \\
         \hline
\end{tabular}
\end{table}

In Fig. \ref{Conductivity} we plot the frequency dependent
conductivity obtained by solving the Maxwell equation numerically
for $\epsilon=0,~0.1,~0.5$ and $0.99$ with $m^{2}L_{AdS}^{2}=0,~-1$
and $-2$.  We find a gap in the conductivity with the gap frequency
$\omega_{g}$. For the same value of $m^2 L_{AdS}^2$, the gap
frequency $\omega_{g}$ decreases with the increase of the constant
$\epsilon$. In each plot, the real part of the conductivity,
Re[$\sigma$], approaches to a limit when the frequency grows. The
limit for the case $\epsilon=0$ is one, but general it increases as
$\epsilon$ increases.  The imaginary part of conductivity
Im[$\sigma$] becomes zero when $\omega\rightarrow \infty$, but it
goes to infinity when the frequency approaches zero.

In Fig. \ref{WgTc} we present the ratio $\omega_{g}/T_{c}$  as a
function of the balance parameter with fixed values $m^2
L_{AdS}^{2}=0$ $-1$ and $-2$, which shows that the ratio
$\omega_{g}/T_C$ almost linear decreases with the increase of the
balance constant.

From Figs. \ref{Conductivity}, \ref{WgTc} and table
\ref{ConductivityTc}, we find that the ratio of the gap frequency in
conductivity $\omega_g$ to the critical temperature $T_c$ in this
black hole reduces to Cai's result $\omega_g/T_c\approx 13$
\cite{Cai-Zhang} found in the Ho\v{r}ava-Lifshitz black hole with
the  condition of the detailed balance for small $\epsilon$, while
it tends to the Horowitz-Roberts relation $\omega_g/T_c\approx 8$ as
$\epsilon \rightarrow 1$.  Our result provides a bridge between the
two kind of the relations.

\section{conclusions}

The behavior of the holographic superconductors in the
Ho\v{r}ava-Lifshitz gravity has been investigated in this manuscript
by introducing a complex scalar field and a Maxwell field in a
planar black-hole background. We first present a detailed analysis
of the condensation of the operator $\mathcal {O}_{+}$ by the
numerical and analytical methods for the Ho\v{r}ava-Lifshitz black
hole without the condition of the detailed balance. We obtain
consistent analytical result with that obtained numerically by
selecting the appropriate matching point. It is found that, as the
parameter of the detailed balance $\epsilon$ increases with fixed
mass of the scalar field $m$, the condensation gap becomes smaller,
corresponding to the larger critical temperature, which means that
the scalar hair can be formed easier for the larger $\epsilon$. And
it is shown that, for the same $\epsilon $, the condensation gap
becomes larger if $m^2$ becomes less negative, which means that it
is harder for the scalar hair to form as the mass of the scalar
field becomes larger. We then studied the electrical conductivity in
the Ho\v{r}ava-Lifshitz black-hole background and find that the
ratio of the gap frequency in conductivity to the critical
temperature, $\omega_{g}/T_c$, almost linear decreases with the
increase of the balance constant.  For $\epsilon= 0$ the ratio
reduces to Cai's result $\omega_g/T_c\approx 13$ \cite{Cai-Zhang}
found in a Ho\v{r}ava-Lifshitz black hole with the condition of the
detailed balance, while as $\epsilon \rightarrow 1$ it tends to the
Horowitz-Roberts relation $\omega_g/T_c\approx 8$
\cite{HorowitzPRD78}  obtained in the AdS Schwarzschild black hole.
It is interesting to note that our result provides a bridge between
the results for the H\v{o}rava-Lifshitz theory with the condition of
the detailed balance and Einstein's gravity.

\begin{acknowledgments}
This work was supported by the National Natural Science Foundation
of China under Grant No 10875040; a key project of the National
Natural Science Foundation of China under Grant No 10935013; the
National Basic Research of China under Grant No. 2010CB833004, the
Hunan Provincial Natural Science Foundation of China under Grant No.
08JJ3010,  PCSIRT under Grant No IRT0964, and the Construct Program
of the National Key Discipline. S. B. Chen's work was partially
supported by the National Natural Science Foundation of China under
Grant No 10875041 and the construct program of key disciplines in
Hunan Province.

\end{acknowledgments}

\end{document}